\documentclass[pra,twocolumn,showpacs,amsmath,eqsecnum,amssymb,superscriptaddress]{revtex4-1}

\usepackage{graphicx,bm,color,mathptmx,hyperref} 

\newcommand{\op}[1]{\hat{#1}}

\newcommand{\var}[1]{ \mathrm{V} ( #1 )}
\newcommand{\cov}[2]{\mathrm{C} ( #1,  #2) }

\begin{document}

\title{Extremal states for photon number and
  quadratures as gauges for nonclassicality}

\author{Z.~Hradil}
\affiliation{Department of Optics, Palack\'y University,
17. listopadu 12, 771 46 Olomouc, Czech Republic}

\author{J.~\v{R}eh\'{a}\v{c}ek}
\affiliation{Department of Optics, Palack\'y University,
17. listopadu 12, 771 46 Olomouc, Czech Republic}

\author{P. de la Hoz}
\affiliation{Departamento de \'Optica, Facultad de F\'{\i}sica,
Universidad Complutense, 28040~Madrid, Spain}

\author{G.~Leuchs}
\affiliation{Max-Planck-Institut f\"ur die Physik des Lichts, 
G\"{u}nther-Scharowsky-Stra{\ss}e 1, Bau 24, 
91058 Erlangen, Germany}
\affiliation{Department f\"{u}r Physik, 
Universit\"{a}t Erlangen-N\"{u}rnberg, Staudtstra{\ss}e 7, 
Bau 2, 91058 Erlangen, Germany}

\author{L.~L.~S\'{a}nchez-Soto} 
\affiliation{Departamento de \'Optica, Facultad de F\'{\i}sica,
Universidad Complutense, 28040~Madrid, Spain}
\affiliation{Max-Planck-Institut f\"ur die Physik des Lichts, 
G\"{u}nther-Scharowsky-Stra{\ss}e 1, Bau 24, 
91058 Erlangen, Germany}
\affiliation{Department f\"{u}r Physik, 
Universit\"{a}t Erlangen-N\"{u}rnberg, Staudtstra{\ss}e 7, 
Bau 2, 91058 Erlangen, Germany}

\begin{abstract}
  Rotated quadratures carry the phase-dependent information of the
  electromagnetic field, so they are somehow conjugate to the photon
  number.  We analyze this noncanonical pair, finding an exact
  uncertatinty relation, as well as a couple of weaker inequalities
  obtained by relaxing some restrictions of the problem.  We also
  find the intelligent states saturating that relation and complete
  their characterization by considering extra constraints on the
  second-order moments of the variables involved.  Using these
  moments, we construct performance measures tailored to diagnose
  photon-added and Schr\"odinger catlike states, among others.
\end{abstract}
 

\maketitle

\section{Introduction}

Leaving aside interpretational
issues~\cite{Caves:2002ai,Spekkens:2007rz,Pusey:2012lq}, the quantum
state is an essential ingredient of quantum theory: once it is known,
the probabilities of the outcomes of any possible measurement may be
predicted.  Unfortunately, this elusive object cannot be directly
determined and must be inferred from a suitable set of measurements,
which constitutes the province of quantum
tomography~\cite{lnp:2004uq}.  Although this might superficially
appear as a mere statistical estimation, the positivity of the
reconstructed state imposes stringent quantum constraints.
  
Indeed, these constraints endow the set of admissible states with a
rich geometry, the boundaries of which somehow establish the realm of
quantum world.  Delimiting these borders is, in general, a difficult
conundrum.  One way to ease these complications is to look at the
problem using moments of the relevant quantities, with the hope that
only a few of them will be important enough.  As a simple yet
illustrative example of this, let us examine a single-mode quantum
field, which will serve as a thread in this work. The complex
amplitude $\op{a}$ and the photon number
$\op{n} = \op{a}^{\dagger} \op{a}$ for this system must obey
$\langle \op{n} \rangle \ge |\langle \op{a} \rangle |^2$, which can be
readily obtained by a simple application of the Cauchy-Schwarz
inequality~\cite{Titulaer:1965la}.  The extremal states (i.e., those
saturating the inequality) turn out to be coherent states.  Hence, the
difference between the average photon number and the square of the
absolute value of the complex amplitude, which must be always
positive, can be taken, up to second order, as a reliable indicator
of the ``quality'' of a coherent state.

In classical signal processing, intensity and phase are the basic
magnitudes specifying the field. At the quantum level, they translate
into photon number and phase. However, the definition of a
\textit{bona fide} phase operator is beset of difficulties that have
been the object of a heated debate~\cite{Carruthers:1968qf,
  Lynch:1995xr,Barnett:1997hb,Perinova:1998kq,Luis:2000vn}. Here, we
choose a surrogate approach that considers the phase properly encoded
in the field quadratures, as it is routinely done in the theory of
quantum nondemolition measurements~\cite{Imoto:1985dw,
  Levenson:1986wc,Hofmann:2000gb}.  While photon number lies at the
heart of the discrete-variable quantum information, quadratures are
the primary tool in the continuous-variable domain. Photon number and
quadratures bridge these two complementary worlds in a natural way.

Extremal states for these variables were investigated some years ago,
fuelled by the search for noise minimum
states~\cite{Hradil:1990ff,Hradil:1991pb}. More recently, the quite
similar question of the uncertainty relation for the number and the
annihilation operator was addressed ~\cite{Urizar-Lanz:2010zl}.  Our
aim here is to push this research further and explore how these
extremal states can  be used for the diagnostics of nonclassicality.

The plan of the  paper is as follows.  In Sec.~\ref{sec:unnx} we revisit
the uncertainty relations for photon number and rotated quadratures,
as well as loose approximations thereof. Section~\ref{sec:add} rounds
off this discussion by looking at the extra restrictions that quantum
theory imposes on the second-order moments of those variables and
looking at the properties of intelligent states, which obey the
equality in the previous uncertainty relations. Based on those states,
we tailor performance measures  especially germane to verify
photon-added and Schr\"{o}dinger catlike states, among
others. Finally, our conclusions are summarized in Sec.~\ref{sec:con}.
 
 \section{Uncertainty relations for photon number and quadratures}
 \label{sec:unnx}

\subsection{Tight uncertainty relations}

The system we are interested in is a single-mode electromagnetic
field, which can be formally deemed as a harmonic oscillator.
Classically, it is characterized by a complex amplitude that contains
information about both the magnitude and the phase of the field. In
the quantum formalism, the mode is specified by the action of
annihilation ($\op{a}$) and creation ($\op{a}^{\dagger}$) operators
satisfying the basic bosonic commutation relation~\cite{Cohen:2006bh}
\begin{equation}
  \label{eq:ccr}
  [ \op{a}, \op{a}^{\dagger} ] = \op{\openone} \, .
\end{equation}

At optical frequencies, the common way of measuring the field is with
homodyne detection~\cite{Leonhardt:2005dz}. The  readout in this
scheme involves moments of the rotated quadratures
\begin{equation}
  \label{eq:rotquad}
  \op{x}_{\theta} = \frac{1}{\sqrt{2}} 
  (\op{a} \, e^{+i \theta} + \op{a}^{\dagger} \, e^{-i\theta}) \, , 
  \qquad 
  \op{p}_{\theta}=\frac{1}{\sqrt{2}i} 
  ( \op{a} e^{+i \theta}-\op{a}^{\dagger} e^{-i \theta}) \, ,
\end{equation}
where $\theta$ is the phase of the local oscillator that can be
externally varied. The reader should be careful about comparing
results on quadrature, as there are a variety of
normalizations used in the literature. Notice that
$\op{p}_{\theta} = - \op{x}_{\theta +\pi/2}$ and that, for $\theta=0$,
they reduce to the canonical variables $\op{x}$ and $\op{p}$.   
They satisfy the canonical commutation relation (in units
$\hbar =1$ throughout)
\begin{equation}
  \label{eq:ccrxp}
  [\op{x}_{\theta}, \op{p}_{\theta}] = i \op{\openone} \, .
\end{equation}

Since
$\op{x}_{\theta}^{2} + \op{p}_{\theta}^{2} = \op{n} +
\op{\openone}/2$,
where $\op{n} = \op{a}^{\dagger} \op{a}$ is the number operator,
precise knowledge of the eigenvalue of $\op{n}$ restricts the possible
knowledge about the quadratures. This is quantified by the commutation
\begin{equation}
  \label{eq:comrot}
  [ \op{n}, \op{x}_{\theta} ] = -i \op{p}_{\theta} \, ,
\end{equation}
which, in turn, implies the uncertainty relation
\begin{equation}
  \label{eq:uncrot}
  \var{\op{n}}  \  \var{\op{x}_{\theta}}  \ge \frac{1}{4} 
  |\langle  \op{p}_{\theta} \rangle |^{2} \, . 
\end{equation}
Here,
$\var{\op{A}} = \langle \op{A}^{2} \rangle - \langle \op{A}
\rangle^{2}$
denotes the variance of $\op{A}$ and the angular brackets
$\langle \cdot \rangle$ mean averaging over the state of the system
(either pure or mixed).

\begin{figure}
  \includegraphics[width=0.70\columnwidth]{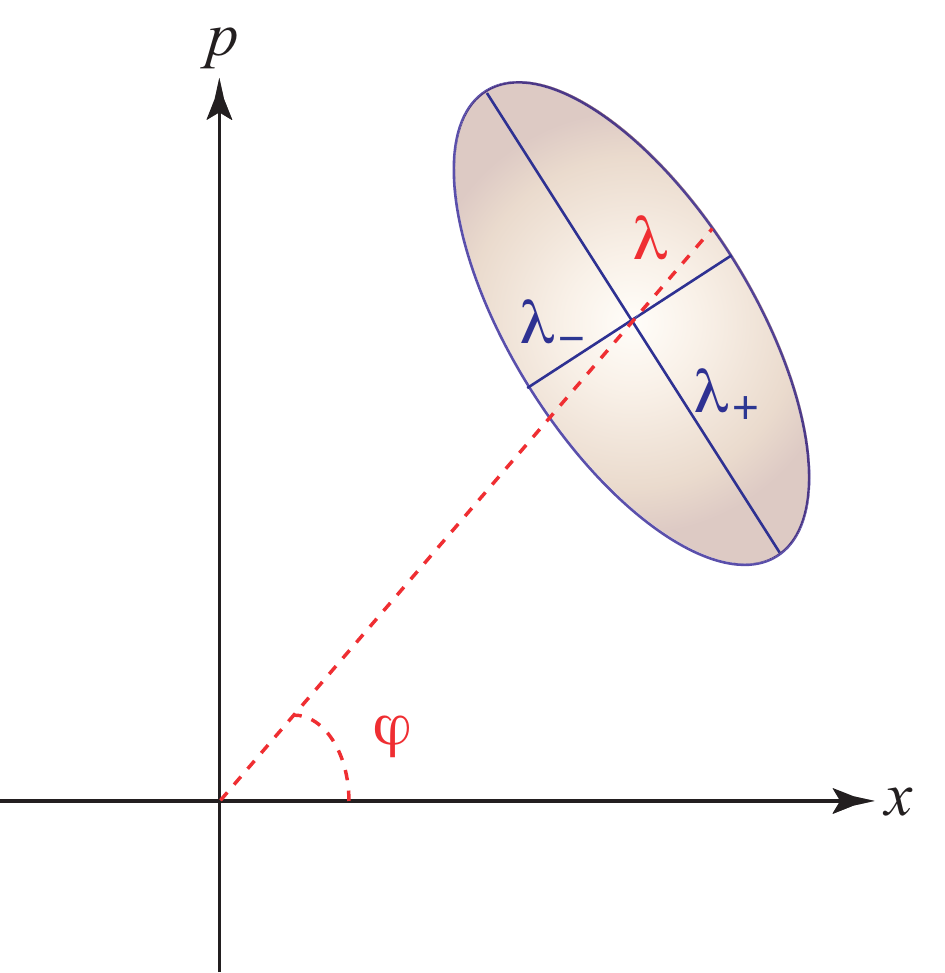}
  \caption{(Color online)  Ball-and-stick diagram in phase-space. 
    $\langle  a \rangle$ is the complex amplitude of the
    field and $\lambda_{\pm}$ are the semiaxes of the uncertainty
    ellipse. The variable $\lambda (\varphi)$ has been defined in
    Eq.~(\ref{eq:lambdagen}).} 
  \label{fig:phasespa}
\end{figure}

Equation~(\ref{eq:uncrot}) is an exact relation, but depends on the
local oscillator phase. Differentiation with respect to $\theta$ leads
to the extremal values $\lambda_{\pm}$ of $\var{\op{x}_{\theta}}$; they can
be written as~\cite{Luks:1988jh}
\begin{equation}
  \label{eq:lambdaeig}
  \lambda_{\pm}^2  =     \cov{\op{a}^{\dagger}}{\op{a}} \pm  
  | \var{\op{a}}|   \, ,
\end{equation} 
where the (symmetrized) covariance is
$ \cov{\op{A}}{\op{B}} = \langle \{ \op{A}, \op{B} \} /2 \rangle -
\langle \op{A} \rangle \langle \op{B} \rangle$.
It is convenient to introduce the quantity $\lambda$ by
\begin{equation}
  \label{eq:lambdagen}
  \lambda^2(\varphi)  =  \lambda_{+}^{2}  \sin^2 \varphi +  
  \lambda_{-}^{2} \cos^2 \varphi \, , 
\end{equation} 
with
$ \varphi = \arg [\var{\op{a}}/2] - \arg [ \langle \op{a} \rangle] $,
in terms of which Eq.~\eqref{eq:uncrot} takes the form
\begin{equation}
  \label{eq:uncertight}
  \var{\op{n}}  \, 
  \left [ \frac{\lambda_{+} \lambda_{-}}{\lambda (\varphi) }  \right ]^{2} 
  \ge |\langle  \op{a} \rangle|^2 \, .
\end{equation}
Apart from the invariant parameters $\lambda_{\pm}$, this expression also
depends on the phase $\varphi$.  However, this alternative presentation will
allow us in the following to devise remarkable
simplifications. Besides, it is closely related to the customary
ball-and-stick representation of quantum states in phase
space~\cite{Bachor:2004hp}, where the quadratures $\op{x}$ and
$\op{p}$ are taken as coordinates.  In this picture, sketched in
Fig.~\ref{fig:phasespa}, the stick corresponds to the average value of
the field $\langle \op{a} \rangle$ and the ball corresponds to the
fluctuations around the mean value. We display this area as a noise
ellipse whose semiaxes are precisely the invariant parameters
$\lambda_{\pm}$. In this way, $\lambda_{\pm}$, which are eigenvalues
of the covariance matrix and related to the universal quantum
invariants~\cite{Dodonov:2000ys}, play a key role in picturing the
noise properties of the state~\cite{Loudon:1989hl,Tanas:1991qr}. The
meaning of $\lambda (\varphi)$ can be gathered at once from that
figure.

As a final remark, we mention that the inequality
(\ref{eq:uncertight}) formally makes it possible to introduce a
quantity like phase variance $\var{\op{\phi}}$ fulfilling a standard
uncertainty relation with $\var{\op{n}}$, namely
\begin{equation}
  \label{eq:putvar}
  \var{\op{\phi}} = \left [ \frac{\lambda_{+} \lambda_{-}}
  {\lambda (\varphi)}  \right ] ^{2} \frac{1}{|\langle  \op{a} \rangle|^{2}} \, .
\end{equation}
Interestingly enough, an explicit calculation shows that this variance
of the putative operator $\op{\phi}$ tallies with the smallest
possible phase resolution in the Shapiro-Wagner phase
concept~\cite{Shapiro:1984ys}, if both quadrature operators are
measured simultaneously.

\subsection{Relaxing the bounds}

The tight uncertainty relation (\ref{eq:uncrot}), or its equivalent
(\ref{eq:uncertight}), convey complete information, but they provide 
phase-dependent bounds. It might be interesting to work out  weaker
inequalities, which are independent of the orientation of the noise
ellipse. 

A first option stems from the trivial observation that, according to
(\ref{eq:lambdagen}), $\inf_{\varphi} \lambda (\phi) =\lambda_{-}$,
so (\ref{eq:uncertight}) can be relaxed to
\begin{equation}
  \label{eq:uncertnt1}
  \var{\op{n}}  \,  \lambda_{+}^{2} \ge |\langle  \op{a} \rangle|^2 \, ,
\end{equation}
or,  using $\cov{\op{a}^{\dagger}}{\op{a}}$,  
\begin{equation}
  \label{eq:uncertnt2}
  \var{\op{n}} \, [ \cov{\op{a}^{\dagger}}{\op{a}} + | \var{\op{a}}| ] \ge
 | \langle \op{a} \rangle |^{2} \, .
\end{equation}

The second possibility comes from the estimate
$[\lambda_{+} \lambda_{-}/\lambda(\varphi) ]^{2} \le \lambda_{+}^{2}
+ \lambda_{-}^{2}$. Now, we can write down
\begin{equation}
  \label{eq:Gturre}
  \var{\op{n}}  \,   (\lambda_{+}^{2} + \lambda_{-}^{2}) \ge 
  |  \langle \op{a} \rangle |^{2} \, ,
\end{equation}
which, using again $\cov{\op{a}^{\dagger}}{\op{a}}$,  
reads as
\begin{equation}
  \label{eq:GTot}
  \var{\op{n}} \,  \cov{\op{a}^{\dagger}}{\op{a}}  \ge 
 |  \langle \op{a} \rangle |^{2} \, .
\end{equation}
This coincides with the expression obtained in
Ref.~\cite{Urizar-Lanz:2010zl}.  In spite of its simplicity, this
inequality has a drawback: it cannot be exactly saturated (yet see the
solution worked out in Ref.~\cite{Adam:2014fk}). This can be confirmed
by noticing that Eq.~(\ref{eq:GTot}) is just the sum of
\begin{equation}
  \label{uSC}
  \var{\op{n}}  \  \var{\op{x}}  \ge \frac{1}{4} |\langle \op{p} \rangle |^2 \, ,
  \qquad
  \var{\op{n}}  \ \var{\op{p}}  \ge \frac{1}{4} |\langle \op{x} \rangle |^2 \, ,
\end{equation}
since
$ | \langle\op{x} \rangle |^{2} + | \langle \op{p} \rangle |^{2} = 2 |
\langle \op{a} \rangle |^{2}$.
But these two relations cannot be saturated
simultaneously~\cite{Rehacek:2008hw,Hradil:2010tw} and, as
consequence, Eq.~(\ref{eq:GTot}) is not tight.

\begin{figure}
  \centerline{\includegraphics[width=\columnwidth]{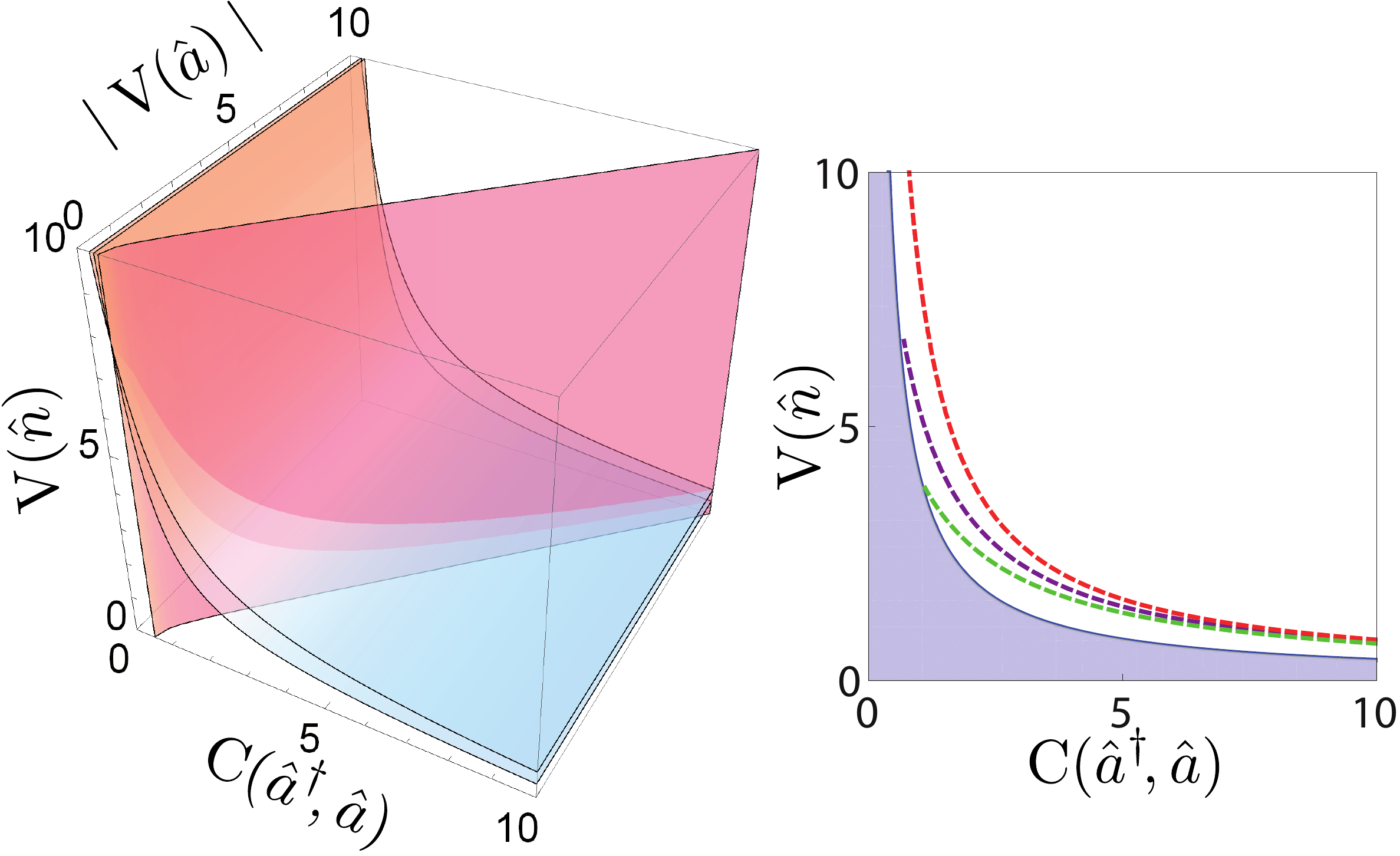}}
  \caption{(Color online) Uncertainty relations (\ref{eq:uncertight})
    and (\ref{eq:uncertnt2}) as a function of the second-order moments
    of the variables involved. The plane along the diagonal
    corresponds to the constraint (\ref{eq:cCov}).  In the right
    panel, we plot sections of the previous figure for several values
    of $| \var{\op{a}}|$ [up to the value permitted by
    (\ref{eq:cCov})]. The continuous line represents the bound
    (\ref{eq:GTot}) and the shaded region designates the forbidden
    states.}
  \label{fig:2}
\end{figure}

In Fig.~\ref{fig:2} we have plotted both approximate inequalities
(\ref{eq:uncertnt2}) and (\ref{eq:GTot}) in a three-dimensional space,
with the moments $\var{\op{a}}$,
$ \cov{\op{a}^{\dagger}}{\op{a}}$, and $\var{\op{n}}$ as axes. The
region above these surfaces are the allowed states.  We have also
plotted several two-dimensional sections for different values of
$| \var{\op{a}}|$. It is evident that (\ref{eq:uncertnt2}) is tighter
than (\ref{eq:GTot}), which actually is independent of
$| \var{\op{a}}|$.

In summary, the exact uncertainty relation (\ref{eq:uncertight}) and
its two weaker approximations (\ref{eq:uncertnt1}) and
(\ref{eq:GTot}), fully specify the complementary nature of photon
number and quadratures. They can be regarded as a sensible alternative
to the more controversial uncertainty relations for number-phase
observables~\cite{Pegg:1989dz,Bialynicki-Birula:1993yk} and their
entropic counterparts~\cite{Rojas-Gonzalez:1995vz,
  Rastegin:2011mk,Rastegin:2012gm}.

\section{Extremal states}
\label{sec:add}

\subsection{Additional restrictions on the moments}

The discussion thus far has capitalized on the variances as the proper
estimator of quantum uncertainties, as it is generally accepted. To
have a complete grasp of the problem, we have to assess also the
constraints arising in the second-order moments involved in the
problem, as they are not independent.

As heralded in the Introduction, an appropriate tool to delimit these
moments is the generalized Cauchy-Schwarz inequality, which can be
jotted down as~\cite{Hradil:1990ff}
\begin{equation}
  \label{eq:GCS}
  | \langle \op{A}^{\dagger}  \op{B} \rangle |^{2} \le 
\langle \op{A}^{\dagger} \op{A} \rangle \,  
\langle \op{B}^{\dagger} \op{B} \rangle  \, .
\end{equation}
The equality occurs only for states where either
$\langle \op{A}^{\dagger} \op{A} \rangle = 0$,
$\langle \op{B}^{\dagger} \op{B} \rangle = 0$, or
$( \op{A} - i r \op{B} ) \op{\varrho} = 0$ for some real scalar
$r$ and $\op{\varrho}$ being the density operator of the state.

A first application of Eq.~(\ref{eq:GCS}), with
$\op{A} = \op{B} = \op{a}$, gives
$\langle \op{a}^{\dagger} \op{a} \rangle \ge \langle \op{a}^{\dagger}
\rangle \langle\op{a}\rangle$.
As a result, from its very definition, the covariance fulfills
\begin{equation}
  \label{eq:cCov}
  \cov{\op{a}^{\dagger}}{\op{a}} \ge \frac{1}{2} \, ,
\end{equation}
which is saturated by the coherent states. Repeating the same
procedure, but now with  $\op{A} = \op{B} = \op{a} - \langle \op{a}
\rangle$, we get
 \begin{equation}
   \label{eq:H}
   | \var{\op{a}} |^{2} \le    {\cov{\op{a}^{\dagger}}{\op{a}}}^{2} -  \frac{1}{4} \, . 
 \end{equation}
This condition is equivalent to require
$ \lambda_+^2 \lambda_{-}^2 \ge 1/4$, which has a direct physical
interpretation in the ball-and-stick diagram analyzed before: for any
physical state, the uncertainty area (in quadrature units) must be
greater or equal than 1/4. For coherent states the noise is equally
distributed in both quadratures,
$\lambda_{-}^{2} = \lambda_{+}^{2} = 1/2$, so they are depicted by a
minimal circle.  In squeezed states, the fluctuations in one
quadrature are reduced below the value $1/2$, at the expense of the
corresponding increased fluctuations in the other quadrature, such
that they preserve the minimum area. Consequently, (\ref{eq:H}) is
saturated by squeezed states.

In this regard,  the condition of squeezing is just $\lambda_{-}^{2}
\le 1/2$, which translates into 
\begin{equation}
   \label{eq:C}
   | \var{\op{a}} | \ge   \cov{\op{a}^{\dagger}}{\op{a}} -   1/2 \, ,
 \end{equation}
which completes (\ref{eq:H}): the states fulfilling this inequality
are not squeezed. 

\begin{figure}
  \centerline{\includegraphics[width=\columnwidth]{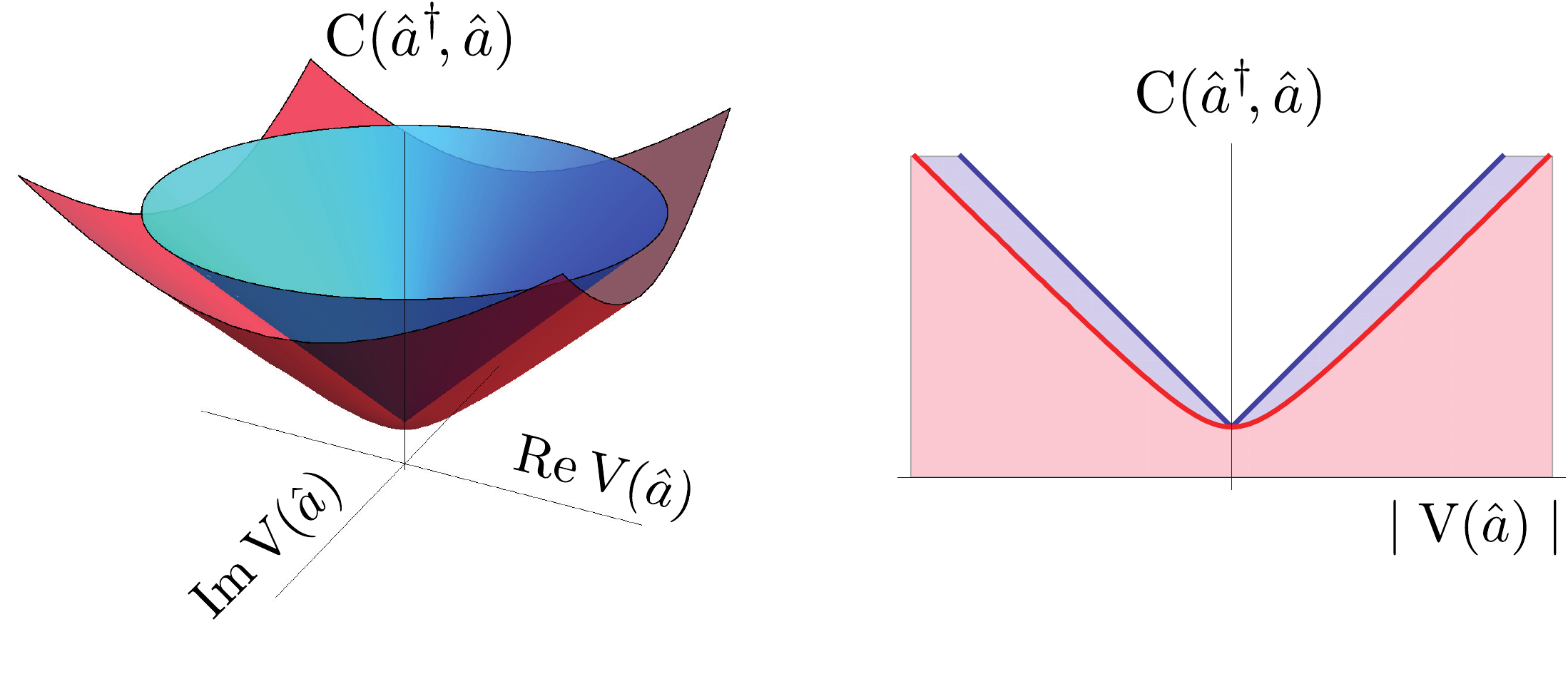}}
  \caption{(Color online) Three-dimensional subspace of all the
    possible second-order moments for a fixed value of $\langle
    \hat{a} \rangle$, ranged by the red hyperboloid. The blue cone is
    the boundary for moment representing squeezed light. In the right,
    we present a section of that figure; the red-shadow region
    represent the feasible states, while the blue-shaded one gives the
    squeezed states.}
  \label{fig:3}
\end{figure}

The inequalities (\ref{eq:H}) and (\ref{eq:C}) can be represented in a
very appealing way if we plot $ \cov{\op{a}^{\dagger}}{\op{a}} $
as a function of the real and imaginary parts of $\var{\op{a}}$, as it
is done  in Fig.~\ref{fig:3}. In these variables, the equality in
(\ref{eq:H}) defines an hyperboloid with vertex in the point $(0, 0,
1/2)$ and all the moments about that hyperboloid are then possible. 
On the other hand, (\ref{eq:C}) defines a cone with vertex in the same
point $(0, 0, 1/2)$: all points below the cone are squeezed. 

Finally, we use once more Eq.~\eqref{eq:GCS}, with
$\op{A} = \op{B} = \op{a}^{2}$, to get
$\langle \op{a}^{\dagger 2} \op{a}^{2} \rangle \ge \langle
\op{a}^{\dagger 2} \rangle \langle \op{a}^{2} \rangle$.
Assuming further the condition of zero complex amplitude,
$\langle \op{a} \rangle = 0$, we have
\begin{equation}
  \label{eq:D}
  \cov{\op{a}^{\dagger 2}}{\op{a}^2} \geq 2 
  \langle\op{a}^\dagger\op{a}\rangle +1 \, ,
\end{equation}
which is saturated by the states spanned on the Hilbert subspace of
the superposition of coherent states $|\pm \alpha\rangle$, a general
solution of the eigenvalue problem
$a^2 | \psi \rangle = \alpha^2 |\psi \rangle$.

\subsection{Intelligent states}
\label{sec:ext}

We have been using the term extremal to loosely refer to those states
for which the inequalities analyzed so far hold as equalities.

If the lower bound in an uncertainty relation is state dependent,
states satisfying the equality in the uncertainty relation need not
give a minimum in the uncertainty product.  This is the case with our
fundamental relation (\ref{eq:comrot}), so it requires a distinction
between intelligent states~\cite{Aragone:1974sf} and minimum
uncertainty product states~\cite{Pegg:2005bf}.

The intelligent states are solutions of the non-Hermitian eigenvalue
problem~\cite{Jackiw:1968ix}
\begin{equation}
  \label{eq:crescent}
  ( \op{n} - i r \, \op{x}_{\theta}  ) | \Psi_{r} \rangle = 
  \Omega | \Psi_{r} \rangle \, , \qquad r \in \mathbb{R} \, , 
\end{equation}
where $\Omega$ is the eigenvalue.  Although the solution to this
equation has been already discussed in Ref. \cite{Hradil:1991pb}, we
provide here a simplified alternative derivation.  By introducing the
complex parameter $\alpha = - i r /\sqrt{2} \, \exp (- i \theta)$,
where $\theta$ is the phase of the quadrature $\op{x}_{\theta}$,
\eqref{eq:crescent} reads
\begin{equation}
  (  \op{a}^{\dagger}   + \alpha^{\ast}  )  ( \op{a} -  \alpha )    
  | \Psi_{\alpha} \rangle = 
  ( \Omega - |\alpha|^{2} ) | \Psi_{\alpha} \rangle \, .
\end{equation}
Since
$ [\op{a} -\alpha, (\op{a}^{\dagger} +\alpha^{\ast} )^M ] = M
(\op{a}^{\dagger} +\alpha^{\ast} )^{M-1}$
for every integer $M$, one quickly guesses that the intelligent states
we are looking for are
\begin{equation}
  \label{crescent_states}
  | \Psi_{\alpha} \rangle = \mathcal{N}  ( 
  a^{\dagger}      + \alpha^{\ast}     )^{M} | \alpha \rangle \, ,
\end{equation}
where $\mathcal{N}$ is a normalization constant and $|\alpha \rangle$
is a coherent state.  We can also expand this expression in the Fock
basis: using the generating function of the generalized Laguerre
polynomials $L_{m}^{a} (x) $~\cite{Morse:1953fk}
\begin{equation}
  (1+t)^{M}  e^{-xt}  = \sum_{n=0}^{\infty} 
  \frac{t^n}{\Gamma(1+M)} L^{M-n}_{n} (x) \, ,  
\end{equation}
we get
\begin{equation}
  \label{crescent_states}
  | \Psi_{\alpha} \rangle =  \mathcal{N}
  \exp (- |\alpha |^{2}/2) \sum_{n=0}^{\infty}
  \frac{\sqrt{n!}}{ M!} (\alpha)^{\ast M-n } L_{n}^{M-n}(- |\alpha |^2) \,  | n  \rangle \, .
\end{equation}
These states were found in a different context by
Yuen~\cite{Yuen:1986ye}, who called them near-photon-number
eigenstates. They are also called crescent states~\cite{Gerry:2005xq}
because the contours of their Wigner function are sheared, due
typically to a Kerr
nonlinearity\cite{Kitagawa:1986xy,Rigas:2013bh}. It is worth stressing
the close similarity of these states, written as in
Eq.~(\ref{crescent_states}), with the expansion for Fock-displaced
states~\cite{Wunsche:1991cs}, although their arguments differ in the
sign, which introduces remarkable differences in the photon-number
distribution.

\begin{figure}
  \centerline{\includegraphics[width=0.65\columnwidth]{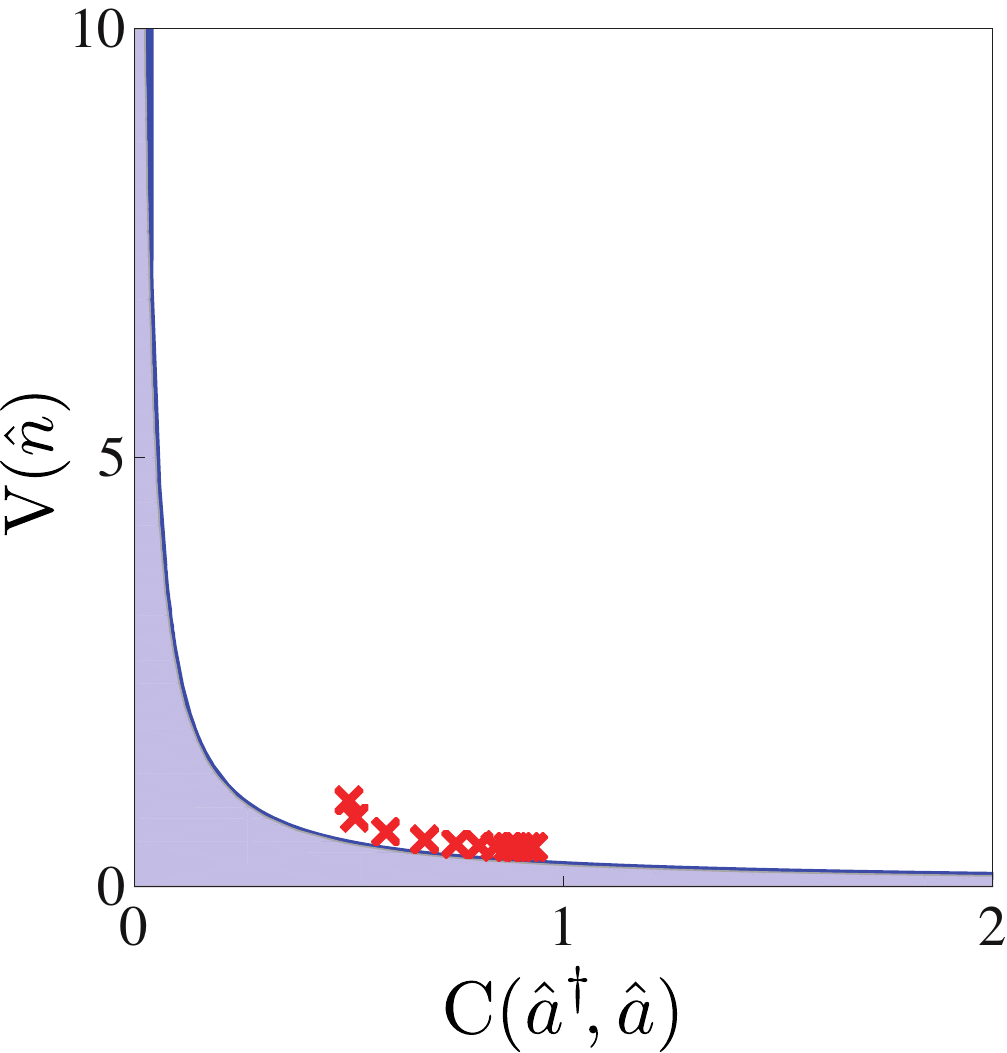}}
  \caption{(Color online) Same plot as in the right panel of
    Fiig.~2. The continous line indicates the weak bound
    (\ref{eq:GTot}), while the crosses represent the approximate
    intelligent states (\ref{eq:sonest}) for several values of
    $\gamma$.}
  \label{fig:4}
\end{figure}

For weak fields $| \alpha | \ll 1 $, the crescent states reduce to the
so-called $M$-photon-added coherent states~\cite{Agarwal:1991pt}
\begin{equation}
  | \Psi_{\alpha} \rangle  \simeq \mathcal{N}  
  \op{a}^{ M \dagger} | \alpha \rangle \,  , 
\end{equation}
while in the strong-field limit $| \alpha | \gg 1 $ they can be well
approximated by the superposition of coherent and single-photon
coherent added states
\begin{equation}
\label{eq:sonest}
  | \Psi_{\alpha} \rangle  \simeq \mathcal{N} ( | \alpha \rangle  +
  \gamma \op{a}^{ \dagger} | \alpha \rangle) \, ,
\end{equation}
with $\gamma = M/\alpha^{\ast}$. For this particular case, we get
\begin{equation}
\langle \op{a} \rangle  =  \alpha + 
\mathcal{N} (\gamma + | \gamma|^{2} \alpha) \, , 
\quad
\langle \op{n} \rangle  = 1 + |\alpha|^{2} + 
\mathcal{N} ( | \gamma |^{2} |\alpha |^{2} - 1) \, , 
\end{equation} 
and the normalization constant is $
  \mathcal{N}^{-1} = 1 + \gamma^{\ast} \alpha + \gamma \alpha^{\ast} + 
|\gamma|^{2} (1 + | \alpha|^{2} )$.
The second-order moments can be analytically computed, although the
expression is a bit involved and of no interest for our purposes here.
In Fig.~\ref{fig:4} we have plotted $\var{\op{n}}$ versus
$\cov{\op{a}^{\dagger}}{\op{a}}$, as we did in Fig.~\ref{fig:2}, for
these states with varying values of $\gamma$. For comparison, we have
included also the bound imposed by the weak uncertainty relation
(\ref{eq:GTot}). As we can appreciate, the intelligent states are
always very close to that bound.

The final  idea we wish to stress is that all these extremal states
can be used as powerful tools to pinpoint important class of
states. Let us look at the crescent states (or the approximation of
$M$-photon-added coherent states treated before). Since they are
intelligent states for  (\ref{eq:comrot}), the coefficient  
\begin{equation}
  \label{criterion1}
  G_1 :=  \frac{   \var{\op{n}}}{  |\langle  \op{a} \rangle|^2}  \, 
  \left [ \frac{\lambda_{+} \lambda_{-}}
    {\lambda (\varphi)}  \right ]^{2}  \ge 1 \, ,
\end{equation}
quantifies how far is a given state from being intelligent.
This diagnosis is an interesting alternative to the full
tomography, which is usually employed to certify this class of
states~\cite{Zavatta:2004mw}.  Since the reconstruction is performed
in an infinite-dimensional Hilbert space, the dimension of the
reconstruction subspace predetermines  the accuracy of the
result.

The inequality (\ref{eq:uncertight}) becomes trivial in case of zero
amplitude.  In that case, however, the inequality (\ref{eq:D} )
provides a saturable bound. As discussed before, the extremal states
are given by the linear superposition of coherent states
$|\pm \alpha \rangle $, including Schr\"odinger catlike states
$ | \alpha\rangle + |- \alpha\rangle. $ The performance measure 
suitable to check these  states is
\begin{equation}
  \label{criterion2}
  G_2 :=    \frac{\cov{\op{a}^{\dagger 2}}{\op{a}^2}}
 { 2 \langle \op{n}\rangle +1 } =  
 \frac{\var{\op{n}}   + 4 ( \lambda_+^2 -1/2 )(  \lambda_-^2 +1/2) }  
 {\langle \op{n} \rangle }   \ge  1 \, ,
\end{equation}
which again provides a robust and simple alternative to more
sophisticated methods. It is obvious that equivalent measures can be
employed for the other extremal states explored here.

\section{Concluding remarks}
\label{sec:con}

In short, we have formulated a tight uncertainty relation for photon number
and rotated quadratures, which can be considered as a sensible and
timely approach to the canonical pair number-phase.  We
have also constructed intelligent states for this uncertainty
relation, retrieving the well-known crescent states.  This saturable
inequality, along with some other obtained from a systematic
application of the Cauchy-Schwarz inequalities to all the second-order
moments of the variables involved, can serve as a handy toolbox for
nonclassical state diagnosis, an alternative to the more onerous and
laborious quantum tomography.

\begin{acknowledgments}
 Many of the ideas in this paper originated at the Workshop on
 Mathematical Methods of Quantum Tomography at Fields Institute
 (Toronto) in february 2013.  We thank G. T\'oth, H. de Guise, and
 B.-G. Englert for fruitful discussions. Z. H. and  J. R. 
 thank the financial assistance of the
  Technology Agency of the Czech Republic (Grant TE01020229) and the
  Czech Ministry of Industry and Trade (Grant FR-TI1/364). G. L. is
  partially funded by EU FP7 (Grant Q-ESSENCE).  Finally, P.~H. and
  L.~L.~S.~S. acknowledge the support from the Spanish MINECO (Grant
  FIS2011-26786) and UCM-Banco Santander Program (Grant GR3/14).
\end{acknowledgments}


%

\end{document}